\documentclass[final,authoryear,3p,times,twocolumn]{elsarticle}

\usepackage{amssymb}

\journal{Applied Radiation and Isotopes}

\begin{document}

\begin{frontmatter}



\title{Activation cross sections of $\alpha$-particle induced nuclear reactions on hafnium and deuteron induced nuclear reaction on tantalum: production of $^{178}$W/$^{178m}$Ta generator}


\author[1]{F. T\'ark\'anyi}
\author[1]{S. Tak\'acs}
\author[1]{F. Ditr\'oi\corref{*}}
\author[2]{A. Hermanne}
\author[3]{A.V. Ignatyuk}
\author[4]{M.S. Uddin}

\cortext[*]{Corresponding author: ditroi@atomki.hu}

\address[1]{Institute for Nuclear Research, Hungarian Academy of Sciences (ATOMKI),  Debrecen, Hungary}
\address[2]{Cyclotron Laboratory, Vrije Universiteit Brussel (VUB), Brussels, Belgium}
\address[4]{Cyclotron Radioisotope Center (CYRIC), Tohoku University, Sendai, Japan}
\address[3]{Institute of Physics and Power Engineering (IPPE), Obninsk, Russia}

\begin{abstract}
In the frame of a systematic study of charged particle production routes of medically relevant radionuclei, the excitation function for indirect production of $^{178m}$Ta through $^{nat}$Hf($\alpha$,xn)$^{178-178m}$Ta nuclear reaction was measured for the first time up to 40 MeV. In parallel, the side reactions $^{nat}$Hf($\alpha$,x)$^{179,177,176,175}$W, $^{183,182,178g,177,176,175}$Ta, $^{179m,177m,175}$Hf were also assessed. Stacked foil irradiation technique and $\gamma$-ray spectrometry were used. New experimental cross section data for the $^{nat}$Ta(d,xn)$^{178}$W reaction are also reported up to 40 MeV.  The measured excitation functions are compared with the results of the ALICE-IPPE, and EMPIRE nuclear reaction model codes and with the TALYS 1.4 based data in the TENDL-2013 library. The thick target yields were deduced and compared with yields of other charged particle ((p,4n), (d,5n) and ($^3$He,x)) production routes for $^{178}$W.
\end{abstract}


\end{frontmatter}


\section{Introduction}
\label{1}
The short-lived (9.3 min) metastable state of the $^{178}$Ta a radioisotope can be used both for diagnostic (PET studies, total $\beta^+$ decay: 1.24\%) as well as for therapeutic purposes (K$_{\alpha1}$ + K${_\alpha2}$ 60\%) \citep{Lacy,Layne,Nichols2013b,Wilson}. It can be produced from long-lived (21.7 d) $^{178}$W via a $^{178}$W/$^{178}$Ta generator. The used production routes include proton and deuteron induced reactions on tantalum and alpha and $^3$He particle induced reactions on hafnium. In the frame of a coordinated research project of the IAEA the evaluation of cross sections of production routes of several medical radioisotopes, including the so called generator isotopes, is in progress \citep{Nichols2013b}. The production routes for a $^{178}$W/$^{178}$Ta generator were previously compiled (not evaluated) in another IAEA project dealing with the physical characteristics and production methods of cyclotron produced radionuclides \citep{Haji}.
The compilation of the available experimental cross section results showed no satisfactory data set. No cross section data are available for the $^{nat}$Hf($\alpha$,xn)$^{178}$W reaction. Among the possible production routes we also measured cross sections (up to 70 MeV) and made theoretical calculation for the $^{nat}$Ta(p,x)$^{178}$W reaction \citep{Uddin}. We also investigated activation cross sections for deuteron induced reactions on Ta up to 40 MeV, but cross section data for $^{178}$W production were not reported \citep{Hermanne}. In this work we experimentally investigate the excitation function of the  $^{nat}$Hf($\alpha$,xn)$^{178}$W reaction and the accompanying side reactions, and by re-evaluating the spectra obtained in our earlier $^{nat}$Ta(d,x) experiment we report cross sections of the $^{nat}$Ta(d,x)$^{178}$W reaction.
To show the capability of different nuclear reaction codes, the measured excitation functions are compared with the results obtained with ALICE-IPPE, EMPIRE and TALYS 1.4 (data from the TENDL-2013 on-line library) nuclear reaction codes. The thick target yields were deduced and also compared with yields of other charged particle production routes for $^{178}$W.
The Ta is nearly monoisotopic, consists of 99.988\% $^{181}$Ta and only 0.012 \% $^{180}$Ta, therefore under the present uncertainty levels we can use in the text the $^{nat}$Ta and the $^{181}$Ta alternately.

\section{Review of earlier measurements of cross section for production of $^{178}$W}
\label{2}
A summary of the earlier and the present experimental investigations (p, d, $\alpha$-particle) found in the literature is presented in Table 1. The numerical cross section and yield data of the earlier measurements were taken from the original works, from the NRDC Experimental Nuclear Reaction Data (EXFOR) database \citep{EXFOR} and from the early Landolt- Bornstein \citep{Semenov} compilations. 

\begin{table*}[t]
\tiny
\caption{Experimental circumstances of the referred measurements}
\begin{center}
\begin{tabular}{|p{0.6in}|p{0.6in}|p{0.8in}|p{1.in}|p{0.8in}|p{1.in}|p{0.4in}|} \hline 
\textbf{Author} & \textbf{Target} & \textbf{Irradiation} & \textbf{Beam current monitor reaction} & \textbf{Measurement of activity} & \textbf{Reaction, data points } & \textbf{Energy range} \newline \textbf{(MeV)} \\ \hline 
\multicolumn{7}{|c|}{\textbf{${}^{181}$Ta(p,4n)${}^{178}$W}} \\ \hline 
\textbf{Rao et al. \newline }(1963)\textbf{\newline }Exfor: C0402 & ${}^{nat}$Ta\newline 84.3 and 21 mg/cm${}^{2}$ & Synchrotron:\newline stacked foil \newline technique\newline  & $^{63}$Cu(p,n)${}^{63}$Zn\newline ${}^{63}$Cu(p,np)${}^{64}$Cu\newline  & Chemical sep.,\newline $\gamma$-NaI(Tl) & ${}^{nat}$Ta(p,4n)  No. 8 & 30-84 \\ \hline 
\textbf{\citep{Birattari}\newline} (1971) \newline Exfor: B0038 & $^{nat}$Ta \newline 20 $\mu$m & cyclotron: \newline stacked foil \newline technique & Faraday cup & $\gamma$-NaI(Tl) & $^{181}$Ta(p,4n)$^{178}$W, No.8 & 28-44.2 \\ \hline 
\textbf{\citep{Zaitseva1994}} \newline Exfor: A0567\textbf{} & ${}^{nat}$Ta\newline 0.20 g/cm${}^{2}$\newline 0.148 g/cm${}^{2}$ & Linac:\newline Stacked foil technique & Faraday cup\newline ${}^{27}$Al(p,x)${}^{24}$Na & No chemical separation\newline $\gamma$-Ge(Li) & ${}^{ }$${}^{1}$${}^{81}$Ta(p,4n)${}^{1}$${}^{78}$W, No.32 & 28.8-78.8 \\ \hline 
\textbf{\citep{Michel}\newline }(2002)\newline Exfor: O1099 & ${}^{nat}$Ta\newline  & cyclotron:\newline synchrotron\newline stacked foil technique & ${}^{27}$Al(p,x)${}^{24}$Na\newline ${}^{nat}$Cu(p,x)${}^{65}$Zn & No chemical separation\newline $\gamma$-HpGe & ${}^{1}$${}^{81}$Ta(p,4n)${}^{1}$${}^{78}$W, No.24 & 70.8-2580 \\ \hline 
\textbf{\citep{Uddin}\newline}(2004)\textbf{\newline }Exfor: E1981 & ${}^{nat}$Ta\newline 10mm & Cyclotron:\newline stacked foil technique & ${}^{nat}$Cu(p,x)${}^{58}$Co\newline ${}^{27}$Al(p,x)${}^{24}$Na & No chemical separation\newline $\gamma$-HpGe & ${}^{181}$Ta(p,4n)${}^{178}$W, No.12 & 28.6-69.0 \\ \hline 
\textbf{\citep{Titarenko}\newline }(2011)\newline Exfor: A0904 & ${}^{nat}$Ta 256-270.9 mg\newline ${}^{181}$Ta 48-59.1 mg & synchrotron & ${}^{27}$Al(p,x)${}^{24}$Na & No chemical\newline separation, g-HpGe & ${}^{nat}$Ta(p,4n)${}^{178}$W, No.10 & 43-1598 \\ \hline 
\multicolumn{7}{|p{1in}|}{\textbf{${}^{181}$Ta(d,5n)${}^{178}$W}} \\ \hline 
\textbf{\citep{Bisplinghoff} \newline }(1974)\newline Exfor A0283 & ${}^{nat}$Ta\newline 20 and 100 mm & Cyclotron:\newline stacked foil technique & ${}^{27}$Al(d,x)${}^{24}$Na & No chemical separation\newline g-Ge(Li) & ${}^{181}$Ta(d,5n)${}^{178}$W, No.14 & 37.8-79.2 \\ \hline 
\textbf{T\'ark\'anyi et al. \newline} (2014)\newline this work\newline Exfor: \textbf{No} & ${}^{nat}$Ta 25 mm & Cyclotron:\newline stacked foil technique  & ${}^{nat}$Ti(d,x)${}^{48}$V\newline ${}^{27}$Al(d,x)${}^{22,24}$Na\newline \newline  & No chemical\newline separation, $\gamma$-HpGe & ${}^{181}$Ta(d,5n)${}^{178}$W, No.8 & 29.6-39.8 \\ \hline 
\multicolumn{7}{|c|}{\textbf{${}^{nat}$Hf(a,xn)${}^{178}$W}} \\ \hline 
\textbf{T\'ark\'anyi et al. \newline }(2014)\newline This work\newline Exfor: No & ${}^{nat}$Hf  \newline 9.62 mm\newline  & cyclotron: stacked foil technique & Faraday cup\newline ${}^{nat}$Ti(a,x)${}^{51}$Cr & No chemical separation\newline $\gamma$-HpGe & ${}^{nat}$Hf(a,xn)${}^{178}$W, No.9 & 14.5-38.9 \\ \hline 
\end{tabular}

\end{center}
\end{table*}

\section{Theoretical calculations}
\label{3}
The measured cross sections were compared with the theoretical effective cross sections calculated by means of three different nuclear reaction computer model codes. For the pre-compound model codes ALICE-IPPE \citep{Dityuk}  and EMPIRE-II \citep{Herman} the parameters for the optical model, level densities and pre-equilibrium contributions were taken as described in \citep{Belgya}. The third set of theoretical values on the figures represent data taken from the  TENDL-2013 online library \citep{Koning2012} calculated with the 1.4 version of TALYS \citep{Koning2007}. The cross sections for isomers in the case of the ALICE code were obtained by using the isomeric ratios calculated with EMPIRE code.

\section{Experimental techniques and data evaluation}
\label{4}

\subsection{The experimental techniques for the $^{nat}$Hf($\alpha$,x) experiment}
\label{4.1}
The experimental method used was similar to that described in our numerous earlier investigations of charged particle induced nuclear reactions for production of medically relevant radioisotopes. Here we report only the most salient features related to reliability of the measured data in table form (see Table 2).
The excitation functions were measured by activation method using the stacked foil technique. Commercial, high purity (Goodfellow) Hf target foils and Ti monitor foils were interleaved in a stack to measure the unknown excitation functions of the radionuclides produced in Hf and to re-measure the well-known excitation function of the monitor reaction in parallel. The thickness and the uniformity of the used targets were determined by weighing the original metal sheets and the individual target foils. The target stack was irradiated at the external beam line of the CGR560 cyclotron of the Vrije Universiteit Brussel (VUB) in a Faraday-cup-like holder to control the beam intensity parameters. 
The primary beam energy was estimated from the parameters of the accelerator and from the extraction, calibrated by TOF method \citep{Sonck}. The excitation function of the simultaneously measured monitor reaction $^{nat}$Ti($\alpha$,x)$^{51}$Cr is shown in Fig. 1 in comparison with the recommended data taken from \citep{TF2001}. An excellent agreement was found after small corrections of the beam intensity (5\% relative to the Faraday cup results) and of the primary beam energy (0.3 MeV).
The decay data were taken from NUDAT 2.6 library \citep{Nudat}, except for a few isotopes (not detailed in NUDAT), for which the decay data are taken from the LBL library (marked in the discussion) \citep{Firestone}. The uncertainty of the energy was increasing from 0.3 MeV for the first foil to 1.5 MeV at the end of the stack. The cross section uncertainties are in the 10-20 \% range and were calculated by using the activation and decay formulas. The uncertainty of cross-sections was derived by summing in quadrature of all individual contributions (beam current (7\%), beam-loss corrections (max. of 1.5\%), target thickness (3 \%), detector efficiency (5\%), photo peak area determination and counting statistics (1-20 \%)). The contributions of the uncertainties of non-linear parameters were neglected (cooling times, half-life). The samples were measured at relatively large distances from the detector surface ($>$ 5 cm) in order to consider them as point-like sources and ensure a better reproducibility.

\begin{table*}[t]
\tiny
\caption{Main experimental parameters and main parameters and methods of the data evaluation of the a-particle induced reactions}
\begin{center}
\begin{tabular}{|p{1.3in}|p{1.4in}|p{1.5in}|p{1.5in}|} \hline 
\multicolumn{2}{|p{1in}|}{\textbf{Main experimental parameters}} & \multicolumn{2}{|p{3.0in}|}{\textbf{Methods of the data evaluation}} \\ \hline 
Incident particle & $\alpha$ (VUB) & $\gamma$-spectra evaluation & Genie 2000, Forgamma \citep{Canberra,Forgamma} \\ \hline 
Method  & Stacked foil & Determination of beam intensity & Faraday cup (preliminary)\newline Fitted monitor reaction (final)\citep{TF1991} \\ \hline 
Target composition & $^{nat}$Hf (9.62 $\mu$m)-target\newline Ti (10.9 $\mu$m)-monitor\newline (repeated 9 times)  & Determination of  primary beam energy & Cyclotron parameters\newline calibrated with TOF \citep{Sonck} \newline Fitted monitor reaction (final) \citep{TF1991} \\ \hline 
Number of Hf  target foils & 9 & Decay data (Table 3) & NUDAT 2.6 \citep{Nudat}\newline and LBNL Isotopes Project \citep{Firestone} \\ \hline 
Accelerator & CGR 560 cyclotron Vrije Universiteit Brussels & Reaction Q-values(Table 3) & Q-value calculator \citep{Q} \\ \hline 
Primary energy & 39.4 MeV & Determination of beam energy\newline in the targets & Andersen (preliminary) \citep{Ziegler}\newline Fitted monitor reaction (final)\newline  \citep{TF1991}  \\ \hline 
Covered energy range & 38.9-14.5 MeV & Uncertainty of energy & Cumulative effects of possible uncertainties\newline (primary energy, target thickness, energy straggling,  correction to monitor reaction) \\ \hline 
Irradiation time & 60  min & Cross sections & Isotopic and elemental cross sections \\ \hline 
Beam current & 100 nA & Uncertainty of cross sections & sum in quadrature of all individual contributions: beam  current  (7\%), beam-loss corrections  (max. 1.5\%),  target thickness  (1\%),  detector efficiency (5\%),  \newline photo peak area determination  and counting statistics   \citep{Error} \\ \hline 
Monitor reaction, [recommended values]  & $^{nat}$Ti(a,x)$^{51}$Cr reaction \citep{TF2001}\newline (re-measured over the whole energy range) & Yield & Physical yield \citep{Bonardi} \\ \hline 
Monitor target and thickness & $^{nat}$Ti, 10.9 $\mu$m &  &  \\ \hline 
detector & HPGe &  &  \\ \hline 
Chemical separation & no &  &  \\ \hline 
$\gamma$-spectra measurements & 3 series &  &  \\ \hline 
Cooling times\newline (and corresponding target-detector distances) & 0.4-2.2 h (20 cm)\newline 3.0-4.5h (15 cm)\newline 415-464 h (5cm)\newline  &  &  \\ \hline 
\end{tabular}

\end{center}
\end{table*}

\begin{table*}[t]
\tiny
\caption{Decay characteristics of the reaction products of the ${}^{ }$${}^{nat}$Hf(a,x)${}^{179,}$${}^{178, }$${}^{177,176,175}$W,${}^{183,}$${}^{182,178g,177,176,175}$Ta, ${}^{179m,177m,175}$Hf  and  ${}^{nat}$Ta(d,xn)${}^{178}$W  investigated reactions and Q-values of react${}^{ }$ions for their productions}
\begin{center}
\begin{tabular}{|p{0.9in}|p{0.6in}|p{0.7in}|p{0.5in}|p{0.8in}|p{0.8in}|} \hline 
\multicolumn{6}{|c|}{\textbf{${}^{nat}$Hf($\alpha$,x)${}^{179,178, 177,176,175}$W,${}^{183.182,178g,177,176,175}$Ta, ${}^{179m,177m,175}$Hf}} \\ \hline 
Nuclide & Half-life & E${}_{\gamma}$(keV) & I${}_{\gamma}$(\%) & Contributing reaction & Q-value\newline (keV) \\ \hline 
\textbf{${}^{1}$${}^{79}$W\newline }$\varepsilon $: 100 \% & 37.05 min & 133.9 & 0.106  & ${}^{176}$Hf($\alpha$,n)\newline ${}^{177}$Hf(a,2n)\newline ${}^{178}$Hf($\alpha$,3n)\newline ${}^{179}$Hf(a,4n)\newline ${}^{180}$Hf(a,5n) & -10927.4\newline -17303.3\newline -24929.3\newline -31028.3\newline -38416.1 \\ \hline 
\textbf{${}^{1}$${}^{78}$W\newline }$\varepsilon $: 100 \%\textbf{} & 21.6 d & The  $\gamma$-lines of the short-lived ${}^{178g}$Ta were used &  & ${}^{176}$Hf($\alpha$,2n)\newline ${}^{177}$Hf($\alpha$,3n)\newline ${}^{178}$Hf($\alpha$,4n)\newline ${}^{179}$Hf($\alpha$,5n)\newline ${}^{180}$Hf($\alpha$,6n) & -17886.8\newline -24262.7\newline -31888.7\newline -37987.7\newline -45375.4 \\ \hline 
\textbf{${}^{1}$${}^{77}$W\newline }$\varepsilon $: 100~\%\textbf{} & 132 min & 115.05\newline 115.65\newline 186.2\newline 186.42\newline 426.98\newline 1036.4 & 8.0 \newline 47 \newline 15.0 \newline 7.3 \newline 12.3 \newline 9.5  & ${}^{174}$Hf($\alpha$,n)\newline ${}^{176}$Hf($\alpha$,3n)\newline ${}^{177}$Hf($\alpha$,4n)\newline ${}^{178}$Hf($\alpha$,5n)\newline ${}^{179}$Hf($\alpha$,6n)\newline ${}^{180}$Hf($\alpha$,7n) & -11791.3\newline -26665.8\newline -33041.7\newline -40667.6\newline -46766.6\newline -54154.4 \\ \hline 
\textbf{${}^{1}$${}^{76}$W\newline }$\varepsilon $: 100~\%\textbf{\newline } & 2.5~h\newline  & 100.20 & 100   & ${}^{174}$Hf(a,2n)\newline ${}^{176}$Hf(a,4n)\newline ${}^{177}$Hf(a,5n)\newline ${}^{178}$Hf(a,6n)\newline ${}^{179}$Hf(a,7n)\newline ${}^{180}$Hf(a,8n) & -18922.8\newline -33797.2\newline -40173.1\newline -47799.1\newline -53898.1\newline -61285.8 \\ \hline 
\textbf{${}^{1}$${}^{75}$W\newline }$\varepsilon $: 100~\%\textbf{\newline } & 35.2 min & 166.69\newline 270.25~\newline  & 9.0\newline 12.6 & ${}^{174}$Hf($\alpha$,3n)\newline ${}^{176}$Hf($\alpha$,5n)\newline ${}^{177}$Hf($\alpha$,6n)\newline ${}^{178}$Hf($\alpha$,7n)\newline ${}^{179}$Hf($\alpha$,8n)\newline ${}^{180}$Hf($\alpha$,9n) & -28002.9\newline -42877.3\newline -49253.2\newline -56879.2\newline -62978.2\newline -70366.0 \\ \hline 
\textbf{${}^{183}$Ta\newline }$\beta $${}^{-}$: 100~\%\textbf{} & 5.1 d & 107.9318\newline 246.0587\newline 353.9904 & 11.0 \newline 27 \newline 11.2  & ${}^{180}$Hf(a,p) & -9349.74 \\ \hline 
\textbf{${}^{182}$Ta\newline }$\beta $${}^{-}$: 100~\%\textbf{\newline } & 114.74 d & 100.10595\newline 1121.290\newline 1189.040\newline 1221.395\newline 1231.004 & 14.20 \newline 35.24 \newline 16.49 \newline 27.23 \newline 11.62  & ${}^{179}$Hf(a,p)\newline ${}^{180}$Hf(a,pn) & -8896.16\newline -16283.92 \\ \hline 
\textbf{${}^{178g}$Ta\newline }$\varepsilon $: 98.76~\%\newline $\beta^{+}$ 1.24\%\newline \textbf{} & 9.31 min & 93.13\newline 1106.09\newline 1340.85\newline 1350.55 & 6.6 \newline 0.54 \newline 1.03 \newline 1.18  & ${}^{176}$Hf($\alpha$,pn)\newline ${}^{177}$Hf($\alpha$,p2n)\newline ${}^{178}$Hf($\alpha$,p3n)\newline ${}^{179}$Hf($\alpha$,p4n)\newline ${}^{180}$Hf($\alpha$,p5n)\newline ${}^{178}$W decay & -16912.8\newline -23288.7\newline -30914.7\newline -37013.7\newline -44401.4 \\ \hline 
\textbf{${}^{178m}$Ta\newline }$\beta^{+}$: 100 \%\newline \textbf{} & 2.36 h & 213.440\newline 325.562\newline 331.613\newline 426.383 & 81.4\newline 94\newline 31.19\newline 97.0 & ${}^{176}$Hf(a,pn)\newline ${}^{177}$Hf($\alpha$,p2n)\newline ${}^{178}$Hf($\alpha$,p3n)\newline ${}^{179}$Hf($\alpha$,p4n)\newline ${}^{180}$Hf($\alpha$,p5n)\newline  & -16912.8\newline -23288.7\newline -30914.7\newline -37013.7\newline -44401.4 \\ \hline 
\textbf{$^{177}$Ta\newline }$\varepsilon $: 100~\%\newline \textbf{} & 56.56 h & 112.9\newline 208.4\newline 745.9\newline 1057.8 & 7.2 \newline 0.94 \newline 0.21 \newline 0.29  & ${}^{174}$Hf($\alpha$,p)\newline ${}^{176}$Hf($\alpha$,p2n)\newline ${}^{177}$Hf($\alpha$,p3n)\newline ${}^{178}$Hf($\alpha$,p4n)\newline ${}^{179}$Hf($\alpha$,p5n)\newline ${}^{180}$Hf($\alpha$,p6n)\newline ${}^{17}$${}^{7}$W decay & -8993.6\newline -23868.17\newline -30244.02\newline -37869.97\newline -43968.96\newline -51356.73\newline  \\ \hline 
\textbf{${}^{176}$Ta\newline }$\varepsilon $: 99.11~\%\newline $\beta^+$: 0.89 \%\textbf{} & 8.09 h & 201.84\newline 710.50\newline 1159.30 & 5.7 \newline 5.4 \newline 24.7  & ${}^{174}$Hf($\alpha$,pn)\newline ${}^{176}$Hf($\alpha$,p3n)\newline ${}^{177}$Hf($\alpha$,p4n)\newline ${}^{178}$Hf($\alpha$,p5n)\newline ${}^{179}$Hf($\alpha$,p6n)\newline ${}^{180}$Hf($\alpha$,p7n)\newline ${}^{17}$${}^{6}$W decay & -17416.7\newline -32291.1\newline -38667.0\newline -46293.0\newline -52391.9\newline -59779.7 \\ \hline 
\textbf{${}^{175}$Ta\newline }$\varepsilon $: 100~\%\textbf{} & 10.5 h & 207.4\newline 266.9\newline 348.5 & 14.0 \newline 10.8 \newline 12.0  & ${}^{174}$Hf($\alpha$,p2n)\newline ${}^{176}$Hf($\alpha$,p4n)\newline ${}^{177}$Hf($\alpha$,p5n)\newline ${}^{178}$Hf($\alpha$,p6n)\newline ${}^{179}$Hf($\alpha$,p7n)\newline ${}^{180}$Hf($\alpha$,p8n)\newline ${}^{17}$${}^{5}$W decay & -24444.7\newline -39319.1\newline -45695.0\newline -53321.0\newline -59420.0\newline -66807.8 \\ \hline 
\textbf{${}^{179m}$Hf\newline }IT: 100~\%\newline 1105.74 keV\textbf{} & 25.05 d & 122.70\newline 146.15\newline 169.78\newline 192.66\newline 236.48\newline 268.85\newline 315.93\newline 362.55\newline 409.72\newline 453.59 & 27.7 \newline 27.1 \newline 19.4 \newline 21.5 \newline 18.8 \newline 11.3 \newline 20.3 \newline 39.6 \newline 21.5 \newline 68  & ${}^{177}$Hf($\alpha$,2p)\newline ${}^{178}$Hf($\alpha$,2pn)\newline ${}^{179}$Hf($\alpha$,2p2n)\newline ${}^{180}$Hf($\alpha$,2p3n) & -14570.71\newline -22196.67\newline -28295.66\newline -35683.43 \\ \hline 
\end{tabular}
\end{center}
\end{table*}

\setcounter{table}{2}
\begin{table*}[t]
\tiny
\caption{Table 3. continued}
\begin{center}
\begin{tabular}{|p{0.9in}|p{0.6in}|p{0.7in}|p{0.5in}|p{0.8in}|p{0.8in}|} \hline 
\multicolumn{6}{|c|}{\textbf{${}^{nat}$Hf($\alpha$,x)${}^{179,178, 177,176,175}$W,${}^{183.182,178g,177,176,175}$Ta, ${}^{179m,177m,175}$Hf}} \\ \hline 
Nuclide & Half-life & E${}_{\gamma}$(keV) & I${}_{\gamma}$(\%) & Contributing reaction & Q-value\newline (keV) \\ \hline

\textbf{${}^{177m2}$Hf\newline }IT: 100~\%\newline 2740.0 keV\newline \textbf{} & 51.4 min & 214.0\newline 277.3\newline 295.1\newline 311.5\newline 326.7\newline 638.2 & 42\newline 78\newline 72\newline 61\newline 68\newline 21.0 & ${}^{176}$Hf($\alpha$,2pn)\newline ${}^{177}$Hf($\alpha$,2p2n)\newline ${}^{178}$Hf($\alpha$,2p3n)\newline ${}^{179}$Hf($\alpha$,2p4n)\newline ${}^{180}$Hf($\alpha$,2p5n) & -21919.75\newline -28295.66\newline -35921.62\newline -42020.62\newline -49408.38 \\ \hline 
\textbf{${}^{175}$Hf\newline }$\varepsilon $: 100~\%\textbf{} & 70 d & 343.40 & 84  & ${}^{17}$${}^{4}$Hf($\alpha$,2pn)\newline ${}^{176}$Hf($\alpha$,2p3n)\newline ${}^{177}$Hf($\alpha$,2p4n)\newline ${}^{178}$Hf($\alpha$,2p5n)\newline ${}^{179}$Hf($\alpha$,2p6n)\newline ${}^{180}$Hf($\alpha$,2p7n)\newline ${}^{175}$Ta decay & -21587.17\newline -36461.59\newline -42837.5\newline -50463.46\newline -56562.46\newline -63950.22 \\ \hline 
\multicolumn{6}{|c|}{\textbf{${}^{nat}$Ta(d,xn)${}^{178}$W}} \\ \hline 
Nuclide & Half-life & E${}_{\gamma}$(keV) & I${}_{\gamma}$(\%) & Contributing reaction & Q-value\newline (keV) \\ \hline 
\textbf{${}^{1}$${}^{78}$W\newline }$\varepsilon $: 100 \%\textbf{} & 21.6 d &  &  & ${}^{180}$Ta(d,4n)\newline ${}^{181}$Ta(d,5n) & -17676.4\newline -25253.1 \\ \hline 
\end{tabular}
\begin{flushleft}
\tiny{\noindent When complex particles are emitted instead of individual protons and neutrons the Q-values have to be decreased by the respective binding energies of the compound particles: np-d: 2.2 MeV; 2np-t: 8.48 MeV; 2p2n-a: 28.30 MeV

\noindent Decrease Q-values for isomeric states with level energy of the isomer

\noindent Abundance of isotopes in natural hafnium: ${}^{174}$Hf (0.16\%), ${}^{176}$Hf (5.26\%), ${}^{177}$Hf (18.60\%), ${}^{178}$Hf (27.28\%), ${}^{179}$Hf (13.31\%), ${}^{180}$Hf (13.31\%)\textbf{}

\noindent Abundance of isotopes in natural tantalum: ${}^{180}$Ta(0.012\%), ${}^{181}$Ta(99.988\%)

}
\end{flushleft}

\end{center}
\end{table*}

\begin{figure}
\includegraphics[width=0.5\textwidth]{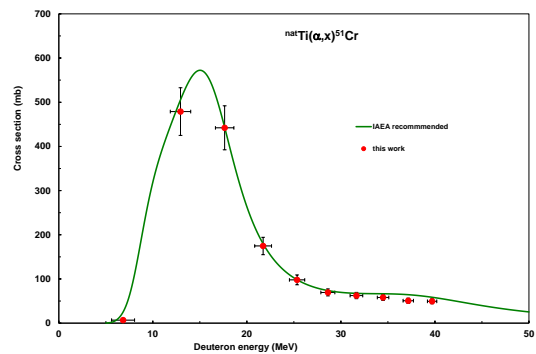}
\caption{Comparison of the simultaneously measured $^{nat}$Ti($\alpha$,x)$^{51}$Cr monitor reaction with the recommended curve }
\label{fig:1}       
\end{figure}

\subsection{The experimental techniques for the $^{nat}$Ta(d,x)$^{178}$W  experiment}
\label{4.2}
The experimental method, origin of decay data and handling of uncertainties are similar as described in the previous paragraph and were reported in detail in our previous work on $^{nat}$Ta(d,x) reactions up to 40 MeV deuteron energy \citep{Hermanne}.

\section{Results}
\label{5}
\subsection{Cross sections}
\label{5.1}

\subsubsection{Cross sections of the $^{nat}$Hf($\alpha$,x)$^{179,177,176,175}$W, $^{183,182,178g,177,176,175}$Ta, $^{179m,177m,175}$Hf reactions}
\label{5.1.1}
The cross sections for all activation products assessed for the $^{nat}$Hf($\alpha$,x) experiment are shown in Figures 2 – 17 and the numerical values are presented in Tables 4-6. The reactions responsible for the production of a given activation product and their Q-values are given in Table 3. The radioisotopes of W are produced via ($\alpha$,xn) reactions, the tantalum radioisotopes by ($\alpha$,pxn) and by EC-$\beta^+$ of W and by $\beta^-$-decay Hf radioisotopes, the Hf radioisotopes are produced by ($\alpha$,2pxn) and EC of Ta.
\vspace{0.3 cm}
\textbf{5.1.1.1 Production of $^{179}$W}\\
The measured cross sections for $^{179}$W ($T_{1/2}$ = 37.05 min) production include the full contribution of the decay of the short-lived isomer ($T_{1/2}$ = 6.40 min, IT: 99.71\%). The experimental and theoretical excitation functions are shown in Fig. 2. In case of ALICE and EMPIRE the total production cross section is given for $^{nat}$Hf($\alpha$,xn)$^{179}$W. The isomeric ratio for the isomer 1/2$^-$ is less than 0.03 and the calculated isomer production is about 30-50 times lower. The values of the theoretical codes are systematically higher, but the agreement between the experimental and theoretical data is more or less acceptable. 

\begin{figure}
\includegraphics[width=0.5\textwidth]{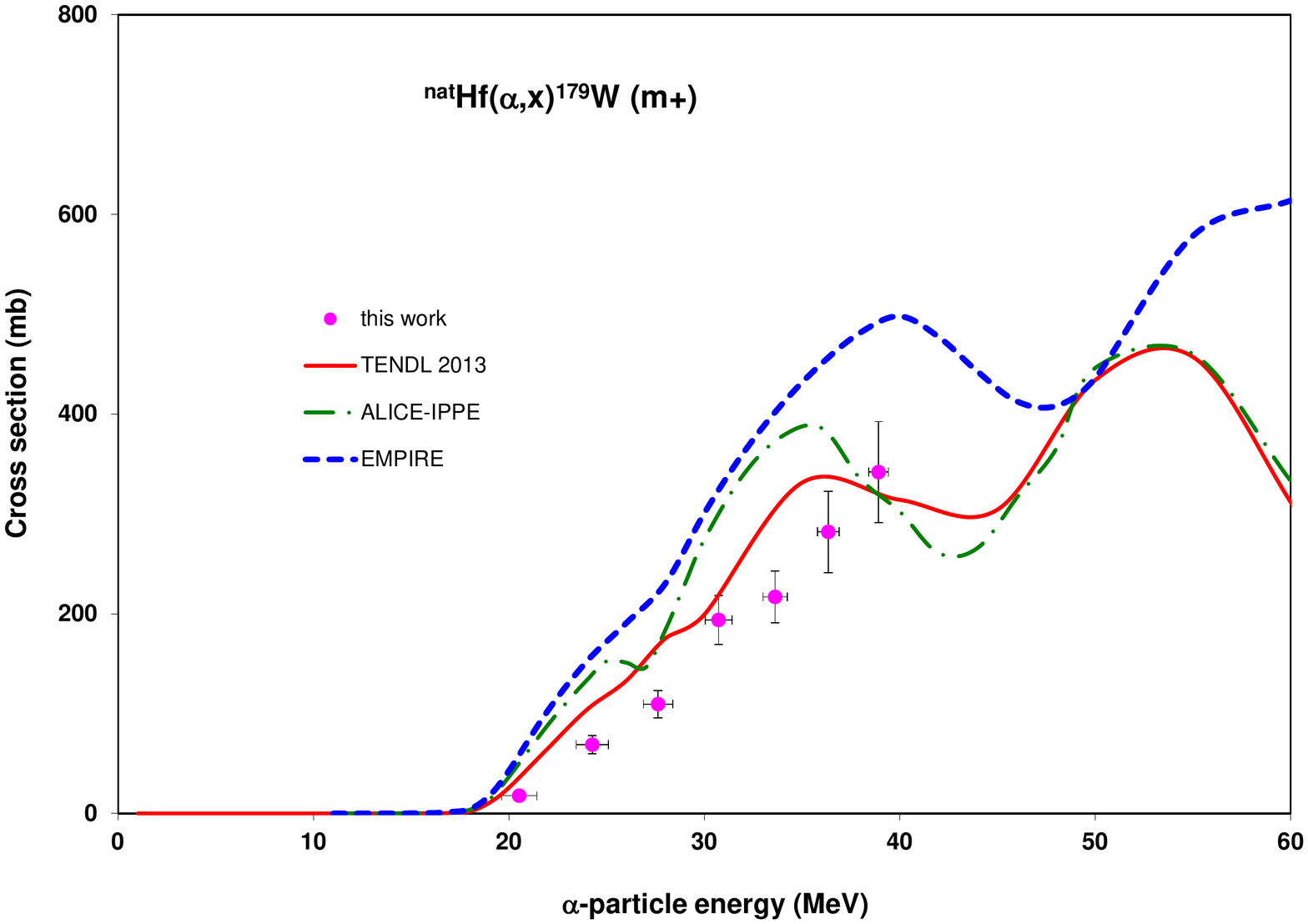}
\caption{Experimental and theoretical excitation functions for $^{nat}$Hf($\alpha$,xn)$^{179}$W}
\label{fig:2}       
\end{figure}


\vspace{0.3 cm}
\textbf{5.1.1.2 Production of $^{178}$W}\\
In the decay of the long-lived $^{178}$W ($T_{1/2}$ = 21.6 d) no measurable $\gamma$-rays are generated. The production cross section of $^{178}$W can however be assessed through $\gamma$-lines of the short-lived $^{178g}$Ta daughter radioisotope ($T_{1/2}$ = 9.31 min) after the complete decay of the directly produced 178gTa ($T_{1/2}$ = 2.45 h). The experimental and model cross sections are shown in Fig. 3. The best approximation is given by the EMPIRE and TALYS codes.

\begin{figure}
\includegraphics[width=0.5\textwidth]{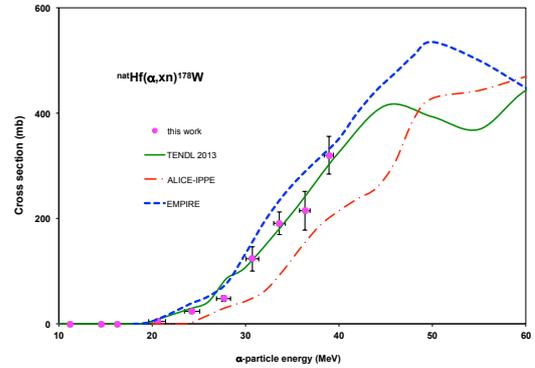}
\caption{Experimental and theoretical excitation functions for $^{nat}$Hf($\alpha$,xn)$^{178}$W}
\label{fig:3}       
\end{figure}

\vspace{0.3 cm}
\textbf{5.1.1.3 Production of $^{177}$W}\\
The measured activation cross sections of $^{177}$W ($T_{1/2}$ = 132 min) are shown in Fig. 4. The agreement with the model calculations ranges from acceptable to good. 

\begin{figure}
\includegraphics[width=0.5\textwidth]{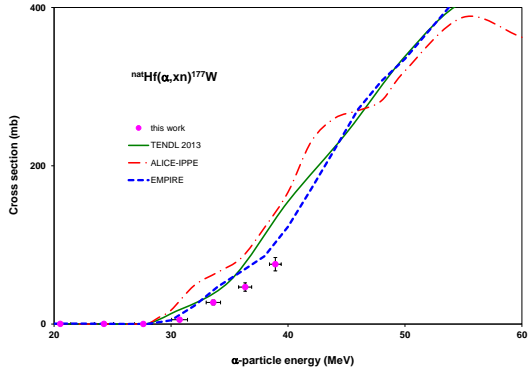}
\caption{Experimental and theoretical excitation functions for $^{nat}$Hf($\alpha$,xn)$^{177}$W}
\label{fig:4}       
\end{figure}

\vspace{0.3 cm}
\textbf{5.1.1.4 Production of $^{176}$W}\\
No datasets were found for the decay characteristics of $^{176}$W ($T_{1/2}$ = 2.5 h) in NUDAT 2.6. The LBL decay database contains only relative intensities for the decay $\gamma$-lines. We calculated cross sections assuming a 100 \% absolute intensity for the most abundant 100.20 keV $\gamma$-line. According to Fig. 5 the magnitudes of experimental and the theoretical cross sections are comparable. In fact, in the investigated energy range production is only seen on the very low abundance $^{174}$Hf, rise starts (according to the theory) by reaction on $^{176}$Hf (threshold: 33 MeV).

\begin{figure}
\includegraphics[width=0.5\textwidth]{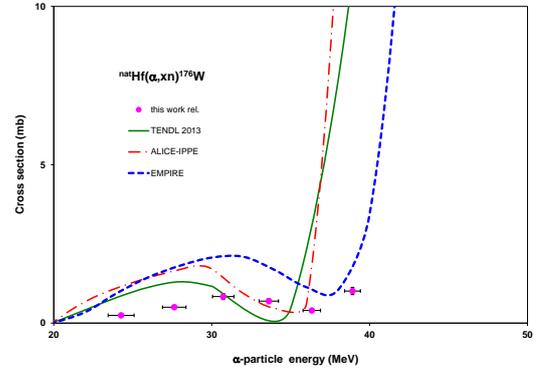}
\caption{Experimental and theoretical excitation functions for $^{nat}$Hf($\alpha$,xn)$^{176}$W}
\label{fig:5}       
\end{figure}

\vspace{0.3 cm}
\textbf{5.1.1.5 Production of $^{175}$W}\\
For $^{175}$W ($T_{1/2}$ = 35.2 min) no decay datasets were found in NUDAT 2.6. The energies and intensities of the $\gamma$-lines were taken from the LBL decay database. The shapes of the experimental and the theoretical data are similar in the overlapping energy range, but the theoretical data are systematically higher (Fig. 6).

\begin{figure}
\includegraphics[width=0.5\textwidth]{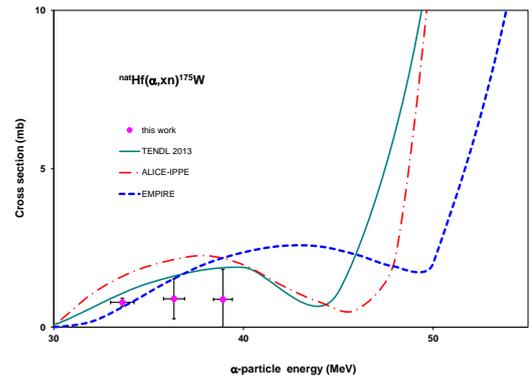}
\caption{Experimental and theoretical excitation functions for $^{nat}$Hf($\alpha$,xn)$^{175}$W}
\label{fig:6}       
\end{figure}

\vspace{0.3 cm}
\textbf{5.1.1.6 Production of $^{183}$Ta}\\
The $^{183}$Ta ($T_{1/2}$ = 5.1 d) can only be produced via the $^{180}$Hf($\alpha$,p) reaction. For  $^{nat}$Hf($\alpha$,x)$^{183}$Ta and $^{nat}$Hf($\alpha$,x)$^{182m}$Ta the EMPIRE code gives too  low  cross  sections  Fig. 7 and 8). It is assumed to be a consequence of a bad pre-equilibrium model for the alpha induced reaction. The ALICE results look quite reasonable for both reactions.

\begin{figure}
\includegraphics[width=0.5\textwidth]{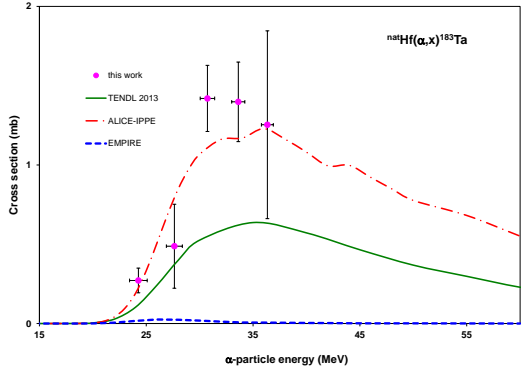}
\caption{Experimental and theoretical excitation functions for $^{nat}$Hf($\alpha$,x)$^{183}$Ta}
\label{fig:7}       
\end{figure}

\vspace{0.3 cm}
\textbf{5.1.1.7 Production of $^{182}$Ta}\\
The measured cross section of $^{182}$Ta ($T_{1/2}$ = 114.74 d) includes the full contribution of the decay of the short-lived isomeric states ($T_{1/2}$ = 15.84 min, IT: 100 \% and 283 ms, IT: 100 \%). According to Fig. 8 the ALICE code gives the best agreement with the experiment.

\begin{figure}
\includegraphics[width=0.5\textwidth]{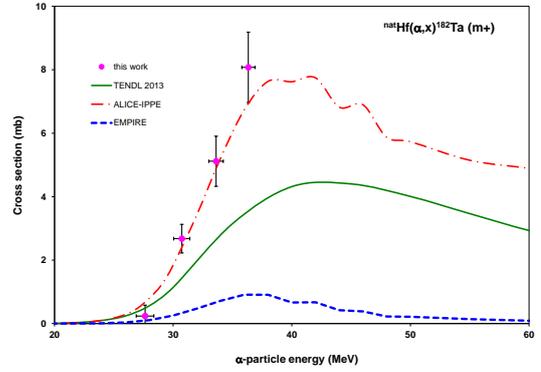}
\caption{Experimental and theoretical excitation functions for $^{nat}$Hf($\alpha$,x)$^{182}$Ta}
\label{fig:8}       
\end{figure}

\vspace{0.3 cm}
\textbf{5.1.1.8 Production of $^{178m}$Ta}\\
For the high spin state of $^{178m}$Ta ($T_{1/2}$ = 2.36 h, I$^\pi$ = 7-) no decay datasets were found in NUDAT 2.6; the data were taken from \citep{Firestone}. Decay of $^{178}$W does not contribute to the population of this sate of $^{178m}$Ta. The experimental data are significantly higher than the predictions of the theoretical codes (Fig. 9).
No experimental data were found for the production of the 9.31 min half-life (1+) $^{178}$Ta ground-state due to the short half-life. The results of our model calculations are shown in Fig. 10.

\begin{figure}
\includegraphics[width=0.5\textwidth]{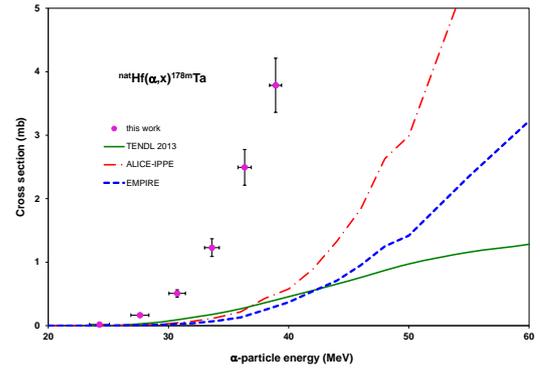}
\caption{Experimental and theoretical excitation functions for $^{nat}$Hf($\alpha$,x)$^{178m}$Ta}
\label{fig:9}       
\end{figure}

\begin{figure}
\includegraphics[width=0.5\textwidth]{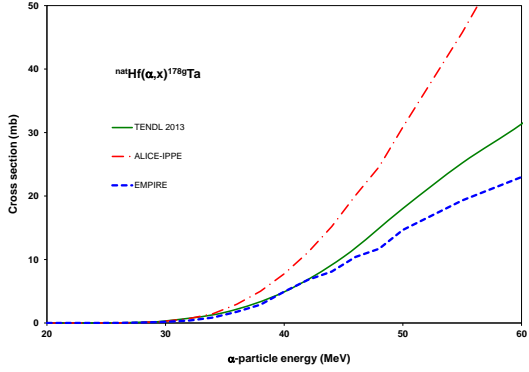}
\caption{Theoretical excitation functions for $^{nat}$Hf($\alpha$,x)$^{178g}$Ta}
\label{fig:10}       
\end{figure}

\vspace{0.3 cm}
\textbf{5.1.1.9 Production of $^{177}$Ta}\\
The cumulative cross sections of $^{177}$Ta ($T_{1/2}$ = 56.56 h) include the direct production and the contribution from the decay of $^{177}$W ($T_{1/2}$ = 132 min, ε: 100 \% $\beta^+$). Cross section is nearly the same as for $^{177}$W. A small contribution from direct reaction exists. The experimental data are in good agreement with the predictions of the different model codes for cumulative production (Fig. 11). Also the TENDL-2013 prediction for the direct production confirms our conclusion from the experiment.

\begin{figure}
\includegraphics[width=0.5\textwidth]{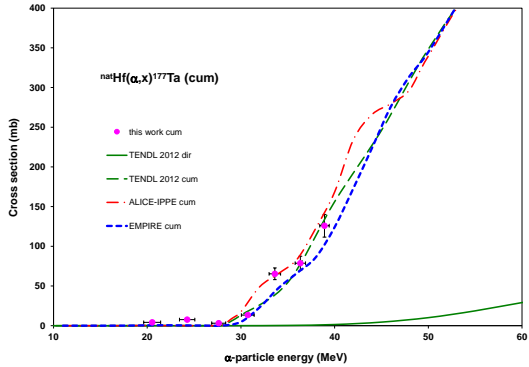}
\caption{Experimental and theoretical excitation functions for $^{nat}$Hf($\alpha$,x)$^{177}$Ta}
\label{fig:11}       
\end{figure}

\vspace{0.3 cm}
\textbf{5.1.1.10 Production of $^{176}$Ta}\\
The cross section data for production of $^{176}$Ta ($T_{1/2}$ = 8.09 h) are cumulative including full contribution of nearly total decay of $^{176}$W ($T_{1/2}$ = 2.5 h). We obtained only a few cross section data near the effective threshold. The direct production is significantly lower, compared to the indirect one (Fig. 12).

\begin{figure}
\includegraphics[width=0.5\textwidth]{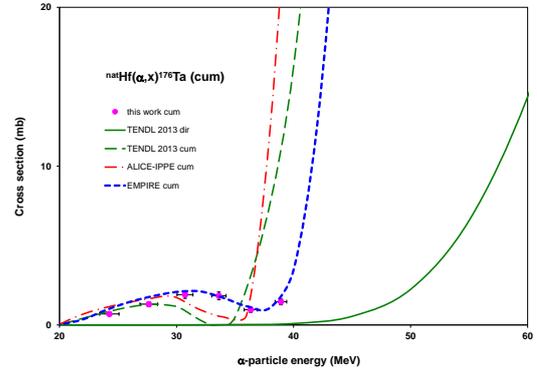}
\caption{Experimental and theoretical excitation functions for $^{nat}$Hf($\alpha$,x)$^{176}$Ta}
\label{fig:12}       
\end{figure}

\vspace{0.3 cm}
\textbf{5.1.1.11 Production of $^{175}$Ta}\\
The measured cumulative cross sections of the $^{175}$Ta ($T_{1/2}$ = 10.5 h) are shown in Fig. 13 in comparison with the theory. The cumulative cross sections were deduced from $\gamma$-spectra measured after nearly complete decay of the $^{175}$W parent isotope ($T_{1/2}$ = 35.2 min).  The three codes reproduce the cumulative experimental excitation function well. According to Fig. 13 the main contribution is indirect.

\begin{figure}
\includegraphics[width=0.5\textwidth]{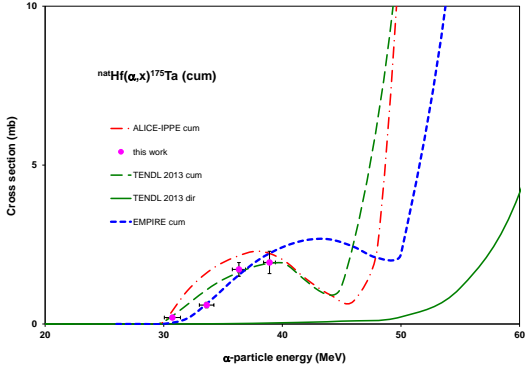}
\caption{Experimental and theoretical excitation functions for $^{nat}$Hf($\alpha$,x)$^{175}$Ta}
\label{fig:13}       
\end{figure}

\vspace{0.3 cm}
\textbf{5.1.1.12 Production of $^{179m}$Hf}\\
In only one spectrum could we identify signal with acceptable statistics from the strong $\gamma$-line at 362 keV, indicating the presence of the long-lived isomeric state of $^{179m}$Hf (25/2)$^-$, $T_{1/2}$ = 25.05 d, IT: 100 \%). The very low corresponding cross section for production via ($\alpha$,2pxn) reactions on higher mass stable Hf isotopes is indicated in Fig. 14.
The magnitude of the predictions of ALICE and EMPIRE are lower compared to the experimental values (see Fig. 14). No theoretical data in the TENDL-2013 library exist. 

\begin{figure}
\includegraphics[width=0.5\textwidth]{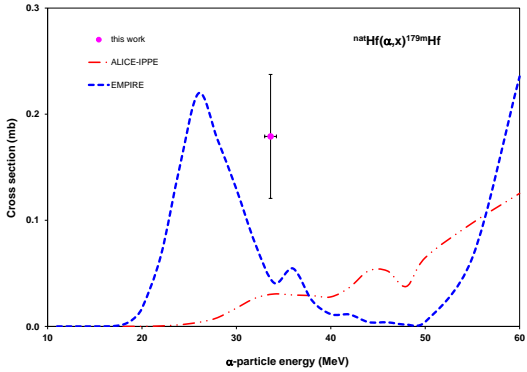}
\caption{Experimental and theoretical excitation functions for $^{nat}$Hf($\alpha$,x)$^{179m}$Ta}
\label{fig:14}       
\end{figure}

\vspace{0.3 cm}
\textbf{5.1.1.13 Production of $^{177m}$Hf}\\
The $^{177m2}$Hf (37/2)$^-$, $T_{1/2}$ = 51.4 min, IT: 100 \%) is produced directly via ($\alpha$,2pxn) reactions. No theoretical data are reported in TENDL-2013 for production of $^{177m2}$Hf.  The experimental data are comparable with the results predicted by the other two model codes (see Fig. 15). 

\begin{figure}
\includegraphics[width=0.5\textwidth]{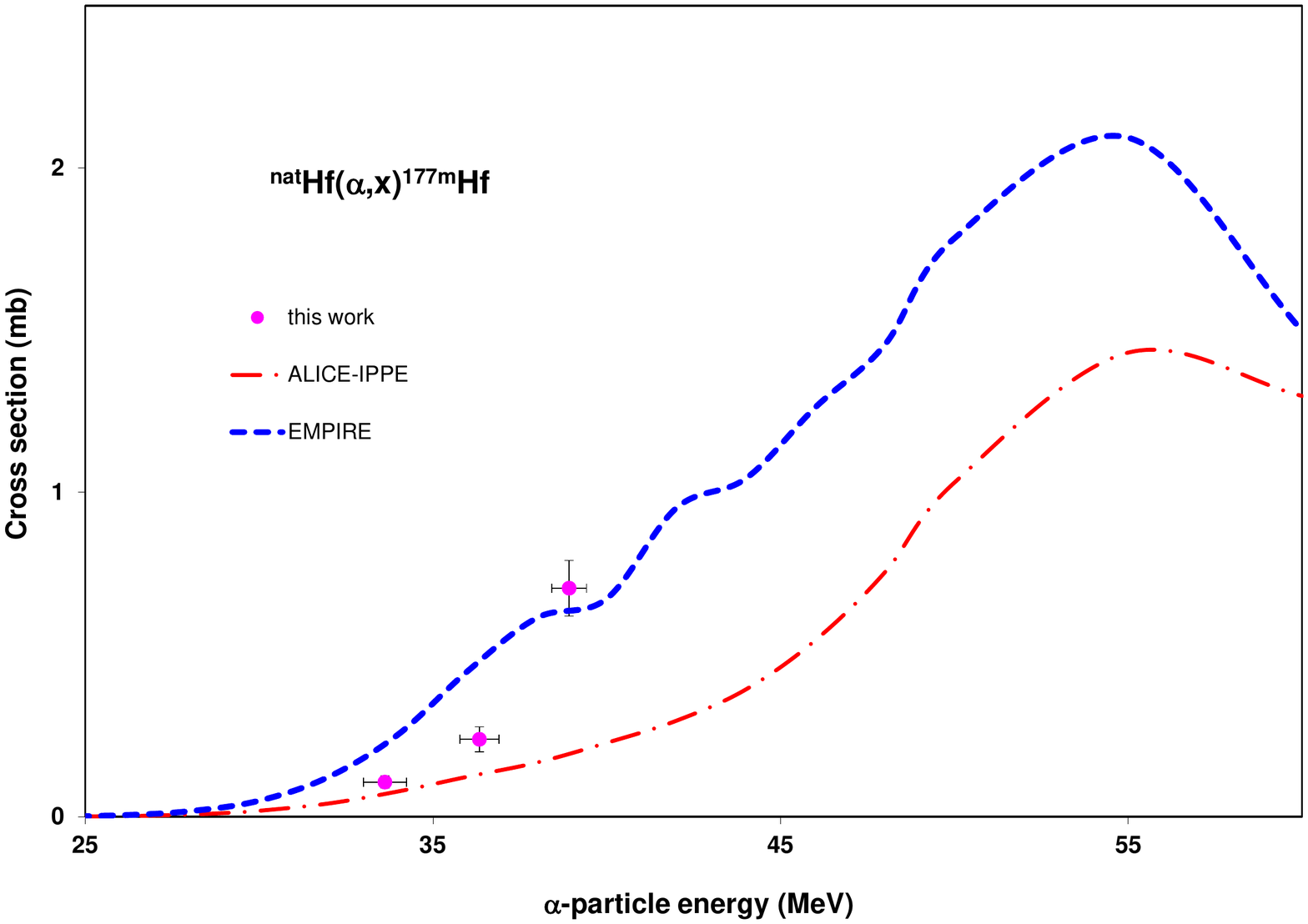}
\caption{Experimental and theoretical excitation functions for $^{nat}$Hf($\alpha$,x)$^{177mTa}$}
\label{fig:15}       
\end{figure}

\vspace{0.3 cm}
\textbf{5.1.1.14 Production of $^{175}$Hf}\\
The cumulative cross sections of $^{175}$Hf ($T_{1/2}$ = 70 d) represent the sum of direct ($\alpha$,2pxn) reactions and indirect production through the decay of the $^{175}$W ($T_{1/2}$ = 35.2 min)-$^{175}$Ta ($T_{1/2}$ = 10.5 h) chain. The results of the model codes are higher compared to the one single measured value; only the EMPIRE code gives acceptable approximation (Fig. 16).

\begin{figure}
\includegraphics[width=0.5\textwidth]{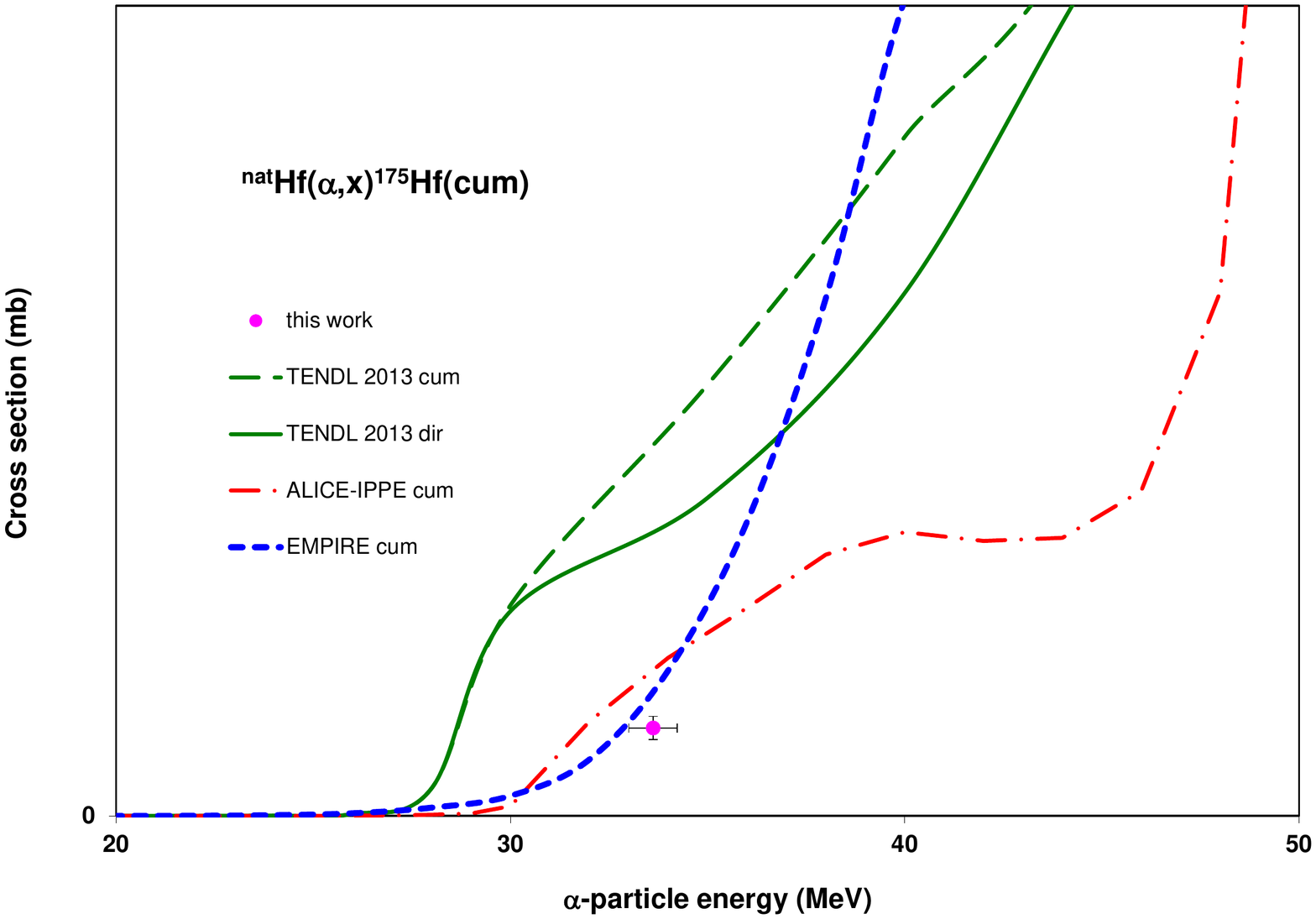}
\caption{Experimental and theoretical excitation functions for $^{nat}$Hf($\alpha$,x)$^{176}$Ta}
\label{fig:16}       
\end{figure}

\begin{table*}[t]
\tiny
\caption{Activation cross sections of tungsten radioisotopes in alpha particle induced reactions on hafnium}
\begin{center}
\begin{tabular}{|p{0.3in}|p{0.1in}|p{0.2in}|p{0.3in}|p{0.1in}|p{0.2in}|p{0.3in}|p{0.1in}|p{0.2in}|p{0.3in}|p{0.1in}|p{0.2in}|p{0.3in}|p{0.1in}|p{0.2in}|p{0.3in}|p{0.1in}|p{0.2in}|} \hline 
\multicolumn{3}{|c|}{\textbf{Energy\newline (MeV)}} & \multicolumn{3}{|c|}{${}^{179}$W\textbf{\newline (mbarn)}} & \multicolumn{3}{|c|}{${}^{178}$W\textbf{\newline (mbarn)}} & \multicolumn{3}{|c|}{${}^{177}$W\textbf{\newline (mbarn)}} & \multicolumn{3}{|c|}{${}^{176}$W\textbf{\newline (mbarn)}} & \multicolumn{3}{|c|}{${}^{175}$W\textbf{\newline (mbarn)}} \\ \hline 
\textbf{38.9} & $\pm$ & \textbf{0.5} & 342 & $\pm$ & 51 & 320 & $\pm$ & 36 & 76 & $\pm$ & 9 & 1.01 & $\pm$ & 0.11 & 0.88 & $\pm$ & 0.88 \\ \hline 
\textbf{36.3} & $\pm$ & \textbf{0.6} & 282 & $\pm$ & 41 & 215 & $\pm$ & 37 & 47 & $\pm$ & 5 & 0.40 & $\pm$ & 0.05 & 0.90 & $\pm$ & 0.63 \\ \hline 
\textbf{33.6} & $\pm$ & \textbf{0.6} & 217 & $\pm$ & 26 & 191 & $\pm$ & 21 & 27 & $\pm$ & 3 & 0.70 & $\pm$ & 0.08 & 0.78 & $\pm$ & 0.13 \\ \hline 
\textbf{30.7} & $\pm$ & \textbf{0.7} & 194 & $\pm$ & 24 & 123 & $\pm$ & 23 & 5.5 & $\pm$ & 0.6 & 0.84 & $\pm$ & 0.10 &  &  &  \\ \hline 
\textbf{27.6} & $\pm$ & \textbf{0.7} & 110 & $\pm$ & 14 & 48 & $\pm$ & 5 & 0.1 & $\pm$ & 0.02 & 0.50 & $\pm$ & 0.06 &  &  &  \\ \hline 
\textbf{24.3} & $\pm$ & \textbf{0.8} & 69 & $\pm$ & 9 & 24 & $\pm$ & 4 & 0.2 & $\pm$ & 0.03 & 0.25 & $\pm$ & 0.03 &  &  &  \\ \hline 
\textbf{20.5} & $\pm$ & \textbf{0.9} & 18 & $\pm$ & 4 & 4.4 & $\pm$ & 0.6 & 0.2 & $\pm$ & 0.03 &  &  &  &  &  &  \\ \hline 
\end{tabular}

\end{center}
\end{table*}

\begin{table*}[t]
\tiny
\caption{Activation cross sections of tantalum radioisotopes in alpha particle induced reactions on hafnium}
\begin{center}
\begin{tabular}{|p{0.3in}|p{0.1in}|p{0.1in}|p{0.3in}|p{0.1in}|p{0.1in}|p{0.3in}|p{0.1in}|p{0.1in}|p{0.3in}|p{0.1in}|p{0.1in}|p{0.3in}|p{0.1in}|p{0.1in}|p{0.3in}|p{0.1in}|p{0.1in}|p{0.3in}|p{0.1in}|p{0.1in}|} \hline 
\multicolumn{3}{|c|}{\textbf{Energy\newline (MeV)}} & \multicolumn{3}{|c|}{\textbf{${}^{183}$Ta\newline (mbarn)}} & \multicolumn{3}{|c|}{\textbf{${}^{182}$Ta\newline (mbarn)}} & \multicolumn{3}{|c|}{\textbf{${}^{178m}$Ta\newline (mbarn)}} & \multicolumn{3}{|c|}{\textbf{${}^{177}$Ta\newline (mbarn)}} & \multicolumn{3}{|c|}{\textbf{${}^{176}$Ta\newline (mbarn)}} & \multicolumn{3}{|c|}{\textbf{${}^{175}$Ta\newline (mbarn)}} \\ \hline 
\textbf{38.9} & $\pm$ & \textbf{0.5} &  & $\pm$ &  &  & $\pm$ &  & 3.8 & $\pm$ & 0.4 & 125.9 & $\pm$ & 14.3 & 1.5 & $\pm$ & 0.2 & 1.9 & $\pm$ & 0.4 \\ \hline 
\textbf{36.3} & $\pm$ & \textbf{0.6} & 1.3 & $\pm$ & 0.6 & 8.1 & $\pm$ & 1.1 & 2.5 & $\pm$ & 0.3 & 78.7 & $\pm$ & 8.9 & 1.0 & $\pm$ & 0.2 & 1.7 & $\pm$ & 0.2 \\ \hline 
\textbf{33.6} & $\pm$ & \textbf{0.6} & 1.4 & $\pm$ & 0.3 & 5.1 & $\pm$ & 0.8 & 1.2 & $\pm$ & 0.1 & 65.3 & $\pm$ & 7.4 & 1.8 & $\pm$ & 0.2 & 0.6 & $\pm$ & 0.1 \\ \hline 
\textbf{30.7} & $\pm$ & \textbf{0.7} & 1.4 & $\pm$ & 0.2 & 2.7 & $\pm$ & 0.5 & 0.51 & $\pm$ & 0.06 & 13.4 & $\pm$ & 1.5 & 1.9 & $\pm$ & 0.2 & 0.2 & $\pm$ & 0.04 \\ \hline 
\textbf{27.6} & $\pm$ & \textbf{0.7} & 0.5 & $\pm$ & 0.3 & 0.2 & $\pm$ & 0.3 & 0.17 & $\pm$ & 0.02 & 3.1 & $\pm$ & 0.4 & 1.3 & $\pm$ & 0.2 &  &  &  \\ \hline 
\textbf{24.3} & $\pm$ & \textbf{0.8} & 0.3 & $\pm$ & 0.1 &  &  &  & 0.02 & $\pm$ & 0.003 & 7.6 & $\pm$ & 0.9 & 0.7 & $\pm$ & 0.1 &  &  &  \\ \hline 
\textbf{20.5} & $\pm$ & \textbf{0.9} &  &  &  &  &  &  &  &  &  & 4.2 & $\pm$ & 0.5 &  &  &  &  &  &  \\ \hline 
\end{tabular}
\end{center}
\end{table*}

\subsubsection{Cross sections of the $^{nat}$Ta(d,xn)$^{178}$W reaction}
\label{5.1.2}
The cross sections for the $^{nat}$Ta(d,xn)$^{178}$W  reaction are shown in Fig. 17 and the numerical values are presented in Table 5. The reactions responsible for the production of a given activation product and their Q-values are given in Table 3. Practically, only the $^{181}$Ta(d,5n) reaction can contribute to the formation of $^{178}$W, as the abundance of  the quasi-stable $^{180}$Ta in $^{nat}$Ta is only 0.012 \%. The earlier reported experimental excitation function of \citep{Bisplinghoff} seems to be shifted to higher energy. The data in the TENDL-2013 library are acceptable, but shifted to lower energies. The best approximation is given by the EMPIRE code in our energy range.

\begin{figure}
\includegraphics[width=0.5\textwidth]{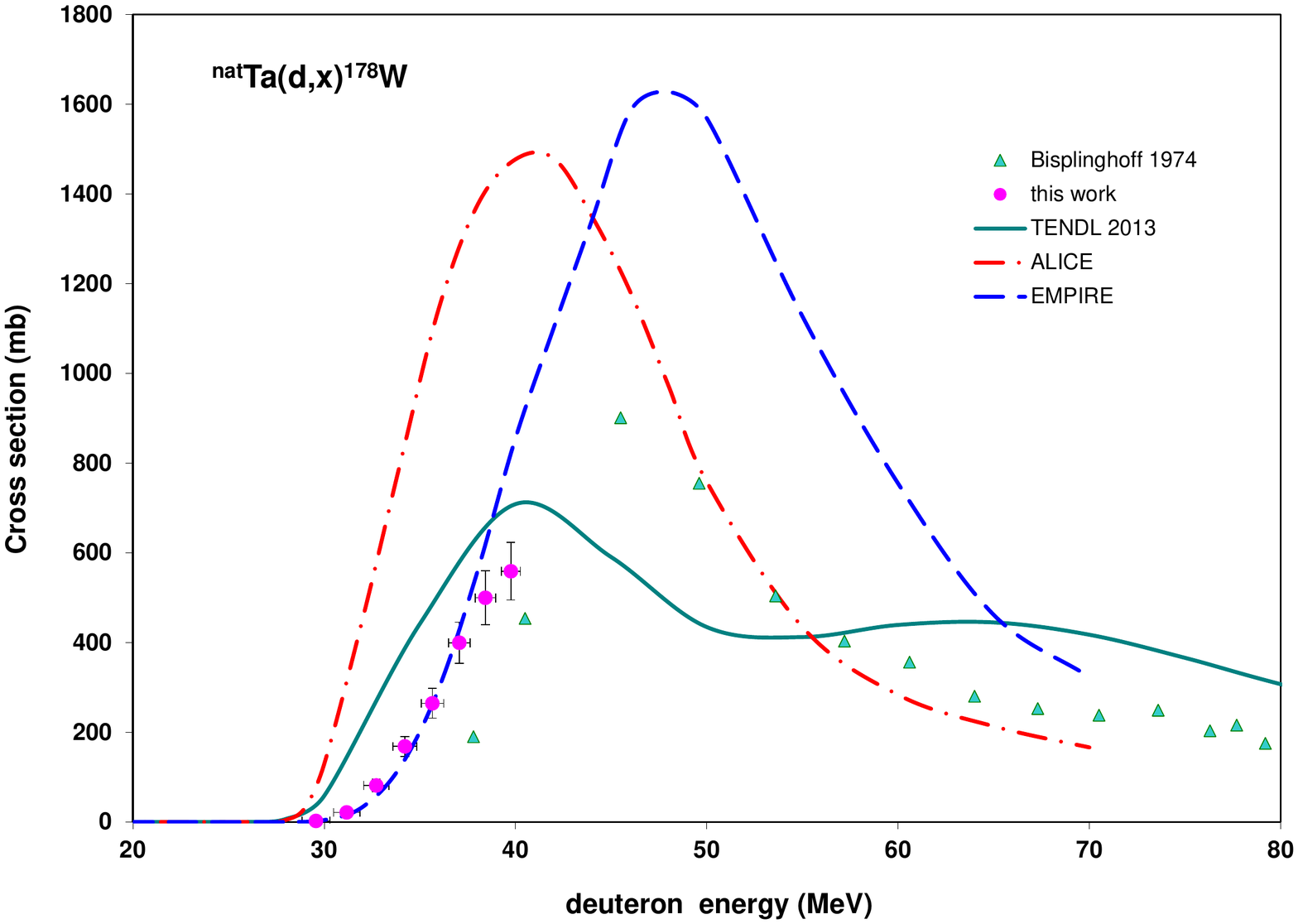}
\caption{Experimental and theoretical excitation functions for $^{181}$Ta(d,5n)}
\label{fig:17}       
\end{figure}

\begin{table*}[t]
\tiny
\caption{Activation cross sections of hafnium radioisotopes in alpha particle induced reactions on hafnium and of the ${}^{181}$Ta(d,5n)${}^{178}$W reaction}
\begin{center}
\begin{tabular}{|p{0.3in}|p{0.1in}|p{0.2in}|p{0.3in}|p{0.1in}|p{0.2in}|p{0.3in}|p{0.1in}|p{0.2in}|p{0.3in}|p{0.1in}|p{0.2in}|p{0.3in}|p{0.1in}|p{0.2in}|p{0.3in}|p{0.1in}|p{0.2in}|} \hline 
\multicolumn{3}{|c|}{\textbf{Energy\newline (MeV)}} & \multicolumn{3}{|c|}{\textbf{${}^{179m}$Hf \newline (mbarn)}} & \multicolumn{3}{|c|}{\textbf{${}^{177m}$Hf\newline (mbarn)}} & \multicolumn{3}{|c|}{\textbf{${}^{175}$Hf\newline (mbarn)}} & \multicolumn{3}{|c|}{\textbf{Energy\newline (MeV)}} & \multicolumn{3}{|c|}{\textbf{${}^{178}$W\newline (mbarn)}} \\ \hline 
\textbf{38.9} & $\pm$ & \textbf{0.5} &  &  &  & 0.70 & $\pm$ & 0.09 &  &  &  & \textbf{39.8} & $\pm$ & \textbf{0.5} & 559 & $\pm$ & 64 \\ \hline 
\textbf{36.3} & $\pm$ & \textbf{0.6} &  &  &  & 0.24 & $\pm$ & 0.04 &  &  &  & \textbf{38.4} & $\pm$ & \textbf{0.5} & 500 & $\pm$ & 60 \\ \hline 
\textbf{33.6} & $\pm$ & \textbf{0.6} & 0.18 & $\pm$ & 0.06 & 0.11 & $\pm$ & 0.02 & 1.1 & $\pm$ & 0.1 & \textbf{37.1} & $\pm$ & \textbf{0.6} & 399 & $\pm$ & 46 \\ \hline 
\textbf{30.7} & $\pm$ & \textbf{0.7} &  &  &  &  &  &  &  &  &  & \textbf{35.7} & $\pm$ & \textbf{0.6} & 264 & $\pm$ & 33 \\ \hline 
\textbf{27.6} & $\pm$ & \textbf{0.7} &  &  &  &  &  &  &  &  &  & \textbf{34.2} & $\pm$ & \textbf{0.6} & 168 & $\pm$ & 22 \\ \hline 
\textbf{24.3} & $\pm$ & \textbf{0.8} &  &  &  &  &  &  &  &  &  & \textbf{32.7} & $\pm$ & \textbf{0.7} & 81 & $\pm$ & 14 \\ \hline 
\textbf{20.5} & $\pm$ & \textbf{0.9} &  &  &  &  &  &  &  &  &  & \textbf{31.2} & $\pm$ & \textbf{0.7} & 21. & $\pm$ & 6 \\ \hline 
\textbf{16.3} & $\pm$ & \textbf{1.0} &  &  &  &  &  &  &  &  &  & \textbf{29.6} & $\pm$ & \textbf{0.7} & 2.0 & $\pm$ & 0.8 \\ \hline 
\end{tabular}

\end{center}
\end{table*}

\subsection{Integral production yields}
\label{5.2}
Thick target yields (integrated yield for a given incident energy down to the reaction threshold) were calculated from fitted curves to our experimental cross section data. The results for physical yields (“production rate”) \citep{Bonardi} are presented in Fig. 18-19. No earlier experimental thick target yield data were found in the literature.

\begin{figure}
\includegraphics[width=0.5\textwidth]{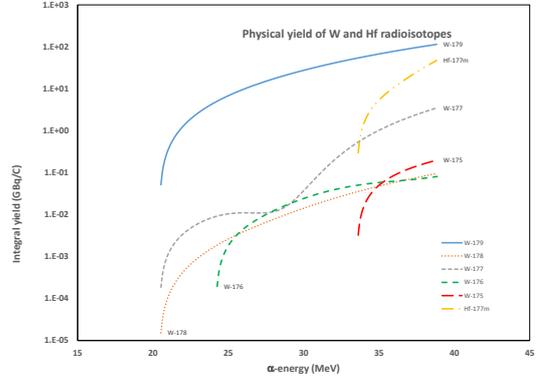}
\caption{Integral yields for production of $^{179,178,177,176,175}$W  and $^{177m}$Hf deduced from the excitation functions}
\label{fig:18}       
\end{figure}

\begin{figure}
\includegraphics[width=0.5\textwidth]{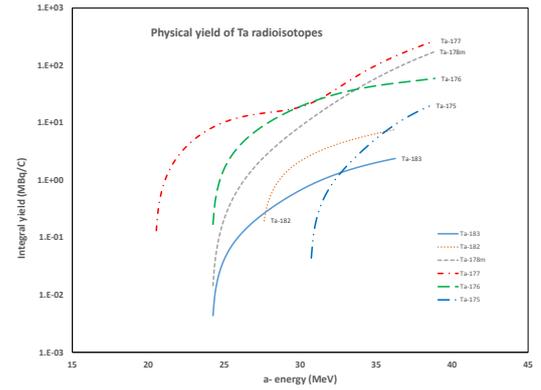}
\caption{Integral yields for production of $^{183,182,178m,177,176,175}$Ta deduced from the excitation functions}
\label{fig:19}       
\end{figure}

\section{Comparison of production routes of $^{178}$W }
\label{6}
The medically interesting radionuclide $^{178}$W (parent in a $^{178}$W/$^{178}$Ta generator) can be produced in various ways: 
\begin{itemize}
\item Proton and deuteron induced spallation–evaporation on heavy mass targets.
\item	Proton induced route via the $^{181}$Ta(p,4n) reaction. 
\item	Deuteron (50 MeV) route via the $^{181}$Ta(d,5n) reaction.
\item	$\alpha$-particle (50-30 MeV incident energy) route via  $^{nat}$Hf($\alpha$,x) and $^{176,177,178}$Hf($\alpha$,xn) reactions.
\item	$^3$He induced route via $^{nat}$Hf($^3$He,x) and $^{176,177,178}$Hf($^3$He,xn) reactions.
\end{itemize}

Among these possibilities the low and medium energy routes (from point of view of the energy ranges of cyclotrons) were investigated also by us (the investigation of the $^3$He process is in progress). While $^{nat}$Ta is practically monoisotopic ($^{181}$Ta-99.988 \%), hafnium has seven stable isotopes, (the isotopic compositions for hafnium and tantalum are shown in Table 2) and optimization of production may ask for enriched targets. In Fig. 23 the integral yield curves were calculated for the most promising reactions for each bombarding particles (p, d, $^3$He, $\alpha$). Each curve was calculated on the basis of the EMPIRE cross-section, taking into account a correction gained by comparison with the experimental data, except in the case of $^3$He, where no experimental data were available. The necessary corrections (based on a comparison with the experimental data) were in the range of 0.65-0.7 in the three corrected cases, so we can suppose that the uncorrected $^3$He-induced yield is overestimated.

\subsection{Spallation-evaporation }
\label{6.1}
The $^{178}$W production using high energy particles was investigated through proton or deuteron induced high energy spallation on tantalum \citep{Michel,Titarenko,Ur}, on mercury \citep{Neuhausen} and on uranium \citep{Casarejos}. The nuclear data and the chemical separation process were investigated by the JINR (Russia group). 

\subsection{$^{181}$Ta(p,4n)$^{178}$W route}
\label{6.2}
Five experimental cross section data sets were found in the literature (see Fig. 20), reported by \citep{Birattari}, \citep{Zaitseva1994},  \citep{Michel} (1 energy point), \citep{Uddin} and \citep{Titarenko} (2 energy points).  The data of the high energy irradiations by Zaitseva and Titerenko show an upward energy shift near the maximum, probably due to the energy uncertainty at the end of a long stack. Moreover the data for the high energy cross section values of Zaitseva seem to be too high and the values at the maximum are too low. Most of routine productions at accelerators were made by using this  reaction \citep{Dmitriev,Pearce}.

\begin{figure}
\includegraphics[width=0.5\textwidth]{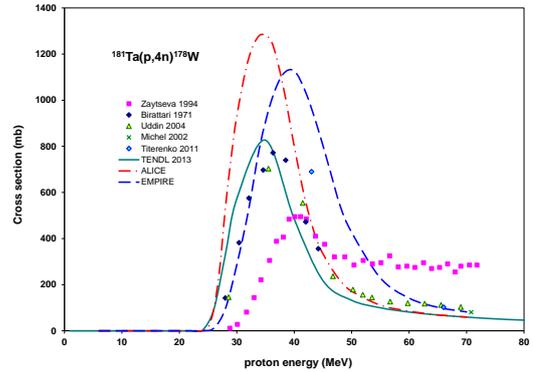}
\caption{Experimental and theoretical excitation functions for $^{181}$Ta(p,4n)$^{178}$W}
\label{fig:20}       
\end{figure}

\subsection{$^{181}$Ta(d,5n)$^{178}$W route}
\label{6.3}
The experimental and theoretical results for the $^{181}$Ta(d,5n) reaction are presented in Fig. 17.  In spite of the large cross sections no routine production was performed using this reaction, due to the high deuteron energy required. By comparing with the $^{181}$Ta(p,4n) reaction, the required incident particle energy is the same, the maximum cross sections are similar, but the yield for deuterons is expected to be lower due to the larger stopping power. Another point is, that in the case of commercial cyclotrons the maximum energy for deuterons is only 50 \% of the maximum energy for protons. No information is available on the real use of deuterons for production of $^{178}$W.

\subsection{176,177,178Hf($\alpha$,xn)$^{178}$W route}
\label{6.4}
According to Fig. 3 the theoretical predictions for the $^{nat}$Hf($\alpha$,x) reaction follow the shape of the experimental data but the values are a little higher. In such a way, on the basis of this agreement we can predict the cross sections of the contributing reactions with high reliability. The theoretical data of contributing reactions from the TENDL-2013, ALICE and EMPIRE are shown in Fig. 21. Only a few routine productions by using alpha particle-induced reaction were already performed \citep{Zaitseva1999}.
As shown in Fig. 21, the maximum of the cross section and consequently the integral yield as a function of the number of emitted neutrons is rising with increasing energy. The comparison of the integral yields with the data on proton and deuteron induced reactions result in the well-known conclusion, that it requires the same energy range, but the yield of alpha induced reactions is lower (Fig. 23). 

\begin{figure}
\includegraphics[width=0.5\textwidth]{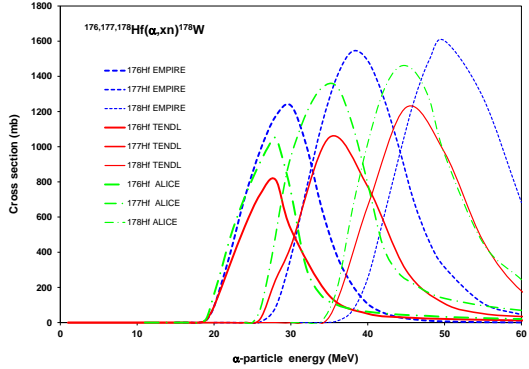}
\caption{Theoretical excitation functions for (ALICE, EMPIRE, TENDL)  $^{176,177,178}$Hf($\alpha$,xn)$^{178}$W}
\label{fig:21}       
\end{figure}

\subsection{$^{176,177,178,179}$Hf($^3$He,xn) route }
\label{6.5}
The theoretical excitation functions (Fig. 22) of the $^3$He-induced reactions show that the cross sections and consequently the yields (Fig.23) are significantly lower compared to the previous reactions, not to mention the high price of the $^3$He gas for the ion source. 

\begin{figure}
\includegraphics[width=0.5\textwidth]{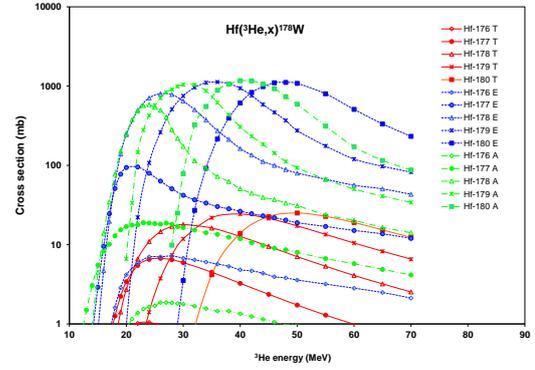}
\caption{Theoretical excitation functions for (ALICE, EMPIRE, TENDL)  $^{176,177,178,179,180}$Hf($^3$He,xn)$^{178}$W}
\label{fig:22}       
\end{figure}

\begin{figure}
\includegraphics[width=0.5\textwidth]{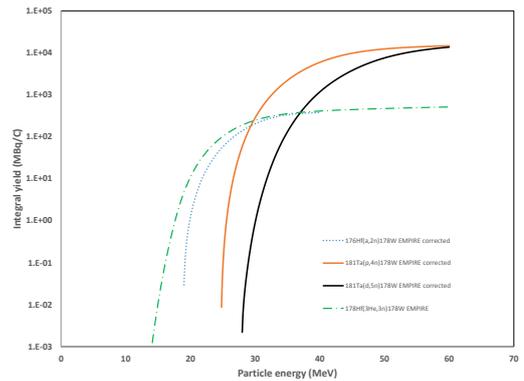}
\caption{Comparison of the integral yield for the $^{178}$W producing proton, deuteron, $^3$He and $\alpha$-induced reactions on different target isotopes}
\label{fig:23}       
\end{figure}

\subsection{Other factors influencing the choice of production route}
\label{6.6}
In the comparison above we compared only the production cross sections and yields. Other important factors should also be discussed, such as target price, target recovery (if necessary), the availability and the price of the bombarding beam, the radionuclide purity, specific activity, etc.:

Spallation requires high energy accelerators, allows use of large mass targets with natural isotopic composition resulting in a large number of radio-products, requiring complicated radiochemical processes for separation and purification. Due to the high energy reactions the target can be placed in the beam dump or tandem targets can be used.

The proton induced reaction on tantalum requires incident energy up to 70 MeV for high yield.  Nowadays 70 MeV, high intensity cyclotrons are to be installed in a few institutes. Cheap natural Ta targets can be used that can withstand high intensity beams. High-purity $^{178}$W production (target material and construction, separation chemistry) was investigated by \citep{Dmitriev1997}.

The optimal energy range for the deuteron route (70-35 MeV) is currently not available for accelerators for routine production. Although the targetry is practically the same as for the proton route, the yield is in any case lower.

Alpha induced reactions on hafnium have only one advantage, namely the $^{176}$Hf($\alpha$,2n) reaction  (30-20 MeV) could be implemented at low energy accelerators (for example at commercially available 30 MeV H$^-$cyclotrons having the alpha option for production of $^{211}$At). The production yield will be lower (compared to the proton and deuteron routes) and the targetry is more complicated as a highly enriched target is needed (price, recovery and target preparation). By using $^{nat}$Hf targets the yield will be about 20 times lower, but the targetry is simpler. Another issue is that the thermal conductivity of the metal hafnium is 2.5 times lower compared to tantalum, and also the maximum beam intensity of the alpha beams is significantly lower than proton intensity at commercial cyclotrons.

The $^3$He-induced reactions are not competitive. They can be used at low energy accelerators but the yield is very low even when using enriched targets. The targetry needed is the same as for $\alpha$-particles. The price of the $^3$He gas for ion source is however high, even by using a $^3$He recovery system. Recently built commercial cyclotrons have no $^3$He options.

\section{Summary and conclusion}
\label{7.}
We present experimental cross sections for the $^{nat}$Hf($\alpha$,x)$^{179,178,177,176,175}$W, $^{183,182,178g,177,176,175}$Ta,  $^{179m,177m,175}$Hf reactions up to 39 MeV $\alpha$-particle energy  for the first time, and new results for the $^{181}$Ta(d,5n) reaction up to  40 MeV. Our excitation functions are correlated to the $^{nat}$Ti($\alpha$,x)$^{51}$Cr and $^{27}$Al(d,x)$^{24}$Na monitor reactions measured simultaneously over the whole covered energy range. The new experimental data are compared with the results of the ALICE-IPPE, EMPIRE and TALYS (TENDL-2013) nuclear reaction model codes, and in the case of $^{181}$Ta(d,5n) reaction also with the results of an earlier measurement in the literature. The theoretical description of experimental data is acceptable by using standard parameters. 
The light ion activation routes of the $^{178}$W for production of the medically related $^{178}$W/$^{178}$Ta generator are compared, showing that in the medium energy range the $^{181}$Ta(p,4n), at low energies the $^{176}$Hf($\alpha$,2n) reactions are the favorite routes.

\section{Acknowledgements}
\label{}
This work was performed in the frame of the HAS-FWO Vlaanderen (Hungary-Belgium) project. The authors acknowledge the support of the research project and of the respective institutions in providing the beam time and experimental facilities.
 



\bibliographystyle{elsarticle-harv}
\bibliography{Hfa}







\end{document}